\documentclass[aps,pra,floatfix]{revtex4}

\usepackage{graphicx}
\usepackage[english]{babel}
\usepackage{amsmath}
\usepackage{amssymb}
\usepackage{slashed}
\usepackage{cancel}
\usepackage{txfonts}

\allowdisplaybreaks

\newcommand{\me}{\mathrm{e}}
\newcommand{\mi}{\mathrm{i}}

\newcommand{\dif}{\mathrm{d}}

\newcommand{\bk}{\mathbf{k}}

\newcommand{\bx}{\mathbf{x}}

\begin{document}
	\title{Finite Temperature Behaviors of $q$-deformed Fermi Gases}
\author{Xun Huang$^{1}$, Xu-Yang Hou$^{1}$, Yan Gong$^{2,3}$, Hao Guo$^{1\ast}$}
\affiliation{$^1$Department of Physics, Southeast University, Nanjing 211189, China}
\affiliation{$^2$School of Material Science and Technology, Beijing Institute of Fashion Technology, Beijing 100029, China}
\affiliation{$^3$Department of Applied Chemistry, Beijing National Laboratory for Molecular Science, College of Chemistry and Molecular Engineering, Peking University, Beijing 100871, China}
\email{guohao.ph@seu.edu.cn}

\begin{abstract}
During the last three decades, non-standard statistics for indistinguishable quantum particles has attracted broad attentions and research interests from many institutions. Among these new types of statistics, the $q$-deformed Bose and Fermi statistics, originated from the study of quantum algebra, are being applied in more and more physical systems. In this paper, we construct a $q$-deformed generalization of the BCS-Leggett theory for ultracold Fermi gases based on our previously constructed $q$-deformed BCS theory. Some interesting features of this $q$-deformed interacting quantum gas are obtained by numerical analysis. For example, in the ordinary Bose-Einstein Condensation regime, the gas presents a fermionic feature instead of bosonic feature if the deformation parameter is tuned suitably, which might be referred to as the $q$-induced ``Bose-Fermi'' crossover. Conversely, a weak sign of the ``Fermi-Bose'' crossover is also found in the ordinary weak fermionic regime.
\end{abstract}
	\maketitle

\section{Introduction}
The bosonic and fermionic superfluidity has long been a hot topic in condensed matter physics,
 and the ultracold Fermi gas with tunable attractive interactions via Feshbach resonance provides a clean and controllable testbed to study superfluidity in the crossover from the fermionic regime, i.e. BCS superfluidity, to the bosonic regime, i.e. Bose-Einstein condensation (BEC)\cite{Son1,KinastPRL05,BruunPRA07,SchaferPRA07,ourlongpaper,ThomasJLTP08,Nascimbene10,EnssPRA12,ThomasPRL14,ThomasScience11,PethickPRL11,ZwierleinNature11,ProchePRA13,SchaeferPRA14,YanPRB14,SchaferPRA15,ThomasPRL15,SchaeferPRL16}. In two-dimensional systems, there exists a new type of indistinguishable quasiparticles obeying the statistics which can be thought of as an interpolating between bosons and fermions. It is called fractional statistics due to the appearing of the braid group instead of the permutation group when identical particles are exchanged upon circling each other. Such quasiparticles are named anyons and have attracted considerable research interests during the last three decades\cite{DunningJSM10,qAnyon93,qAnyon96}.
 An interesting question is, for physical systems of arbitrary spatial dimensions, whether there exists other generalizations to the standard bosons or fermions? The answer is yes! The studies of solutions to the Yang-Baxter equation\cite{Drinfeld,JimboLMP86,JimboBook,qo0} have led to
the concepts of deformed versions of ordinary Lie algebras, which are called quantum algebra in many literatures. The deformed algebra and  deformed superalgebra can be realized by the $q$-deformed bosonic harmonic oscillator and its fermionic counterpart respectively\cite{qb1,qb2,qb3,sqa1,sqa2}, where $q$, usually a positive real number, is the deformation parameter.
These $q$-deformed particles, or simply $q$-particles or quons, obey a deformed commutative or anti-commutative relations similar to anyons, but are substantially different from the latter, since the former can be interpreted as the Fourier components of local field operators whereas the latter are essentially non-local. One must be interested in the experimental realization of $q$-deformed quantum systems. Though there are no direct experimental researches in this field until now, people have already tried to explain some realistic phenomena by $q$-deformed particles. For example, it was argued that photons with a half-integer total angular momentum\cite{hspt} in reduced dimensions may be recognized as $q$-deformed fermions\cite{hspt17}, and the $q$-deformed statistics was also applied to study the emissivity of the light fermionic dark matter in the cooling of the supernova SN1987A\cite{qsn17}.

Many-particle systems consisting of bosons, fermions and anyons have been exhaustively studied. One may also be interested in the properties of many-quon systems\cite{VPJC1,qh11,qfs1,qfs2,qfs3,qfs4,Algin11,QF1,QF2,qfm04}. There have been some studies on the thermodynamics of non-interacting many-$q$-boson and many-$q$-fermion systems. However, due to theoretical complexities, investigations on interacting many-quon systems are still very rare. In nuclear physics, the $q$-deformed nucleon pairs were introduced to study the appearance of possible fermionic condensates\cite{qfnp,qbcs3,qbcs1,qbcs2,qnjl,qnclp01,qnclp03,qnclp03a,qnclp11}. However, it is hard to apply this approximate model to condensed matter systems.
 In Ref.\cite{OurJStat18}, we construct an exact solvable $q$-deformed BCS ($q$BCS) theory by generalizing the temperature Green's function formalism to the two-component many-$q$-fermion system. When $q=1$, it naturally reduces to the ordinary BCS theory. This model possesses some interesting properties. For example, it has a built-in population-imbalanced structure. Thus, even when there is no number polarization between the two species, the system still presents the breached pair state, which only shows in the population-imbalanced case for ordinary interacting Fermi gases.

As pointed out at the beginning, many interacting condensed matter systems, either fermionic or bosonic, can be modulated by ultracold Fermi gases with tunable interactions both experimentally and theoretically. The theoretical foundation of these approaches is the BCS-Leggett theory\cite{Leggett}, in which the BEC state can be realized simply by tuning the inter-particle interaction of an ultracold Fermi gas. Therefore, instead of constructing a theory for interacting $q$-deformed bosonic gas (it is well known that even the many-body theory for ordinary bosons encounter many theoretical difficulties comparing to its fermionic counterpart), we simply generalize our $q$BCS theory to the $q$-deformed version of BCS-Leggett theory, and study the properties of the bosonic state by tuning the inter-particle interaction via the scattering length.

The rest of the paper is organized as follows. In Sec.\ref{s2}, we briefly review the $q$BCS theory, and then generalize it to the $q$-deformed BCS-Leggett theory for ultracold $q$-Fermi gases. The important equations of state will be given there. In Sec.\ref{sc2}, we present our numerical analysis about the finite temperature behaviors of $q$-Fermi gases when adjusting the inter-particle interaction. It can be found that some interesting phenomena show up when the deformation parameter $q$ is tuned to certain values. The conclusion is summarized in the end.

 \section{Theoretical Framework}    \label{s2}

Throughout this paper, we adapt the convention that $\hbar=1$, $k_B=1$.
In general, there are four types $q$-deformed fermi algebra, known as Fermionic Newton (FN), Chaichian-Kulish-Ng (CKN), Parthasarathy-Viswanathan-Chaichian (PVC) and Viswanathan-Parthasarathy-Jagannathan-Chaichian (VPJC) models\cite{Algin11}, which can be converted to one another by certain transformations.
Our model is built upon the VPJC algebra.
Moreover, for a two-component Fermi gas with pseudo spins $\sigma=\uparrow, \downarrow$,
there are two sets of field operators $\psi_{\uparrow,\downarrow}$. It is natural to assume that each field operator has its own deformation parameter. However, to ensure that the algebra is closed under operations between the fields with different spins, the deformed Fermi algebra must be a two-mode generalization of VPJC algebra
\begin{align}\label{qBCSa1}
&\psi_{\bk\uparrow}\psi^\dagger_{\bk'\uparrow}+q\psi^\dagger_{\bk'\uparrow}\psi_{\bk\uparrow}=\delta_{\bk\bk'},\quad
\psi_{\bk\downarrow}\psi^\dagger_{\bk'\downarrow}+q^{-1}\psi^\dagger_{\bk'\downarrow}\psi_{\bk\downarrow}=\delta_{\bk\bk'},\notag\\
&\psi_{\bk\uparrow}\psi_{\bk'\downarrow}+q\psi_{\bk'\downarrow}\psi_{\bk\uparrow}=0,\quad \quad 
\psi^\dag_{\bk\downarrow}\psi^\dag_{\bk'\uparrow}+q\psi^\dag_{\bk'\uparrow}\psi^\dag_{\bk\downarrow}=0,
\end{align}
where $\bar{\uparrow}=\downarrow$ and vice versa.
The first line indicates that the two deformation parameters are the inverse of each another, thus the coupled equations of motion for Green's function and anomalous Green's function form a closed set. For details, please refer to Ref.\cite{OurJStat18}.
At very low temperature, we self-consistently assume that the pairing between $q$-fermions with different spins may occur and lead to the fermionic condensation. In this paper, we only focus on the $s$-wave paring, of which the pairing function is given by $\Delta(\mathbf{x})=g\langle\psi_\uparrow(\mathbf{x})\psi_\downarrow(\mathbf{x})\rangle$ with $g$ being the inter-particle coupling constant. Assuming the chemical potentials for each species are $\mu_{\uparrow,\downarrow}$, and the particle mass is $m$, the Hamiltonian of the $q$BCS model is generalized to take the form\cite{OurJStat18}
\begin{equation}\label{HqBCS}
H=\sum_{\bk,\sigma}\xi_{\bk\sigma}N_{\bk\sigma}+\sum_\bk(\Delta^* S_\bk+\Delta S^\dag_\bk)+\frac{|\Delta|^2}{qg},
\end{equation}
where $\xi_{\bk\sigma}=\frac{\bk^2}{2m}-\mu_\sigma$, $N_{\bk\sigma}$ is the number operator for each species such that
 \begin{align}\label{Na1}
\left[N_{\bk\sigma}, \psi_{\bk'\sigma'}\right]=-\psi_{\bk\sigma'}\delta_{\bk\bk'}\delta_{\sigma\sigma'},\quad
\left[N_{\bk\sigma}, \psi^\dagger_{\bk'\sigma'}\right]=\psi^\dagger_{\bk\sigma'}\delta_{\bk\bk'}\delta_{\sigma\sigma'},
\end{align}
 and
$S_\bk$ and $S^\dagger_\bk$ are spin operators satisfying the algebra
\begin{align}\label{nSa}
&\Big[S^\dag_{\bk},\psi_{\bk'\uparrow}\Big]=\frac{1}{q}\psi^\dag_{-\bk\downarrow}\delta_{\bk\bk'}, \quad \Big[S^\dag_{\bk},\psi_{\bk'\downarrow}\Big]=-\psi^\dag_{-\bk\uparrow}\delta_{\bk\bk'}\notag \\
&\Big[S_{\bk},\psi^\dag_{\bk'\uparrow}\Big]=-\frac{1}{q}\psi_{-\bk\downarrow}\delta_{\bk\bk'},  \quad \Big[S_{\bk},\psi^\dag_{\bk'\downarrow}\Big]=\psi_{-\bk\uparrow}\delta_{\bk\bk'},\notag\\
&\textrm{all other commutators vanish.}
\end{align}
The field operators in the Heisenberg picture are defined as $\psi_\sigma(x)=\me^{H\tau}\psi_\sigma(\bx)\me^{-H\tau}$, $\psi^\dag_\sigma(x)=\me^{H\tau}\psi^\dag_\sigma(\bx)\me^{-H\tau}$ where $\tau$ is the imaginary time $\tau\equiv \mi t$ and $x\equiv (\tau,\bx)$. Thus, the algebras (\ref{Na1}) and (\ref{nSa}) imply the equations of motion for field operators
\begin{align}\label{eomf}
\frac{\partial \psi_\uparrow(x)}{\partial \tau}&=-\left(\frac{(-\mi\nabla)^2}{2m}-\mu_\uparrow \right)\psi_\uparrow(x)+\frac{\Delta(\bx)}{q}\psi^\dagger_\downarrow(x),\notag\\
\frac{\partial \psi^\dagger_\downarrow(x)}{\partial \tau}&=\left(\frac{(-\mi\nabla)^2}{2m}-\mu_\downarrow\right)\psi^\dagger_{\downarrow}(x)+\Delta^\ast(\bx)\psi_{\uparrow}(x).
\end{align}
The single-particle Green's function and anomalous Green's function for $q$-fermions are defined by
\begin{align}\label{GAG}
&G_{\uparrow}(x,x')=-\langle T_{\tau}[\psi_{\uparrow}(x)\psi^{\dag}_{\uparrow}(x')]\rangle=-\langle\psi_{\uparrow}(x)\psi^{\dag}_{\uparrow}(x')\rangle\theta(\tau-\tau')+q\langle\psi^{\dag}_{\uparrow}(x')\psi_{\uparrow}(x)\rangle\theta(\tau'-\tau), \nonumber \\
&G_{\downarrow}(x,x')=-\langle T_{\tau}[\psi_{\downarrow}(x)\psi^{\dag}_{\downarrow}(x')]\rangle=-\langle\psi_{\downarrow}(x)\psi^{\dag}_{\downarrow}(x')\rangle\theta(\tau-\tau')+q^{-1}\langle\psi^{\dag}_{\downarrow}(x')\psi_{\downarrow}(x)\rangle\theta(\tau'-\tau), \nonumber \\
&F_{\uparrow\downarrow}(x,x')=-\langle T_{\tau}[\psi_{\uparrow}(x)\psi_{\downarrow}(x')]\rangle=-\langle\psi_{\uparrow}(x)\psi_{\downarrow}(x')\rangle\theta(\tau-\tau')+q\langle\psi_{\downarrow}(x')\psi_{\uparrow}(x)\rangle\theta(\tau'-\tau), \nonumber \\
&F_{\downarrow\uparrow}(x,x')=-\langle T_{\tau}[\psi_{\downarrow}(x)\psi_{\uparrow}(x')]\rangle=-\langle\psi_{\downarrow}(x)\psi_{\uparrow}(x')\rangle\theta(\tau-\tau')+q^{-1}\langle\psi_{\uparrow}(x')\psi_{\downarrow}(x)\rangle\theta(\tau'-\tau), 
\end{align}
which are consistent with the $q$-algebra (\ref{qBCSa1}).
In most situations, the Green's functions possess both spatial and temporal translational symmetries. Thus, they can be expressed as $G_\sigma(x,x')\equiv G_\sigma(\tau-\tau',\bx-\bx')$ and $F_{\sigma\sigma'}(x,x')\equiv F_{\sigma\sigma'}(\tau-\tau',\bx-\bx')$.
It has been found that these modified Matsubara Green's functions have a special periodic property, i.e. they are in fact scaled by $q$ or $\frac{1}{q}$ when the time duration is translated by a factor $\beta=\frac{1}{T}$. Here $T$ is the temperature and accordingly $\beta$ is the inverse temperature up to a factor of Boltzmann constant. Explicitly, we have
\begin{align}
&G_\uparrow(\tau-\tau',\bx-\bx')=-qG(\tau-\tau'+\beta,\bx-\bx'),\notag\\
&G_\downarrow(\tau-\tau',\bx-\bx')=-\frac{1}{q}G(\tau-\tau'+\beta,\bx-\bx'),\notag\\
&F_{\uparrow\downarrow}(\tau-\tau',\bx-\bx')=-qF_{\uparrow\downarrow}(\tau-\tau'+\beta,\bx-\bx'),\notag\\
&F_{\downarrow\uparrow}(\tau-\tau',\bx-\bx')=-\frac{1}{q}F_{\downarrow\uparrow}(\tau-\tau'+\beta,\bx-\bx').
\end{align}
This property brings an extra $q$-dependent imaginary part to the Matrsbara frequency when performing Fourier transformations to these Green's functions
\begin{eqnarray}\label{mf}
	\omega_n=\frac{(2n+1)\pi}{\beta}\mp\mi\frac{\ln q}{\beta}.
	\end{eqnarray}
where $n$ is an integer, and the signs $\mp$ correspond to $q^{\pm 1}$ respectively. The equations of motion for fields, i.e. Eqs.(\ref{eomf}), lead to two decoupled sets of differential equations for Green's functions, which can be further converted to algebraic equations in the momentum space by performing Fourier transformations. For simplicity, we only present the latter results
\begin{align}\label{dEGAG}
&\left(\mi\omega_n+\frac{\ln q}{\beta}-\xi_{\bk\uparrow}\right)G_{\uparrow}(\mi\omega_n,\bk)+\frac{1}{q}\Delta F^{\dag}_{\uparrow\downarrow}(\mi\omega_n,\bk)=1,\notag  \\
&\left(-\mi\omega_n-\frac{\ln q}{\beta}-\xi_{\bk\downarrow}\right)F^{\dag}_{\uparrow\downarrow}(\mi\omega_n,\bk)-\Delta^{*}G_{\uparrow}(\mi\omega_n,\bk)=0,
\end{align}
and
\begin{align}\label{mEGAG}
&\left(\mi\omega_n-\frac{\ln q}{\beta}-\xi_{\bk\downarrow}\right)G_{\downarrow}(\mi\omega_n,\bk)-\Delta F^{\dag}_{\downarrow\uparrow}(\mi\omega_n,\bk)=1,  \notag\\
&\left(-\mi\omega_n+\frac{\ln q}{\beta}-\xi_{\bk\uparrow}\right)F^{\dag}_{\downarrow\uparrow}(\mi\omega_n,\bk)+\frac{1}{q}\Delta^{*}G_{\downarrow}(\mi\omega_n,\bk)=0.
\end{align}
Here the Matsubara frequency still follows the ordinary definition $\omega_n=\frac{(2n+1)\pi}{\beta}$ to emphasize the dependence of Green's functions on $q$.
Introducing $\mu=\frac{\mu_\uparrow+\mu_\downarrow}{2}$ and $h=\frac{\mu_\uparrow-\mu_\downarrow}{2}$, these equations are solved to give
\begin{align}\label{eud}
 G_{\uparrow}(\mi\omega_n,\bk)&=\frac{|u_{\bk}|^2}{\mi\omega_n+\frac{\ln q}{\beta}-E_{\bk\uparrow}}+\frac{|v_{\bk}|^2}{\mi\omega_n+\frac{\ln q}{\beta}+E_{\bk\downarrow}},  \notag \\
 F^{\dag}_{\uparrow\downarrow}(\mi\omega_n,\bk)&=-\sqrt{q} u_{\bk}^{*}v_{\bk}\left(\frac{1}{\mi\omega_n+\frac{\ln q}{\beta}-E_{\bk\uparrow}}-\frac{1}{\mi\omega_n+\frac{\ln q}{\beta}+E_{\bk\downarrow}}\right),
\end{align}
and
\begin{align}\label{edu}
 G_{\downarrow}(\mi\omega_n,\bk)&=\frac{|u_{\bk}|^2}{\mi\omega_n-\frac{\ln q}{\beta}-E_{\bk\downarrow}}+\frac{|v_{\bk}|^2}{\mi\omega_n-\frac{\ln q}{\beta}+E_{\bk\uparrow}},  \notag\\
\quad \quad F^{\dag}_{\downarrow\uparrow}(\mi\omega_n,\bk)&=\frac{u_{\bk}^{*}v_{\bk}}{\sqrt{q}}\left(\frac{1}{\mi\omega_n-\frac{\ln q}{\beta}-E_{\bk\downarrow}}-\frac{1}{\mi\omega_n-\frac{\ln q}{\beta}+E_{\bk\uparrow}}\right),
\end{align}
where $E_{\bk}=\sqrt{\xi_{\bk}^2+\frac{1}{q}|\Delta|^2}$, $\xi_\bk=\frac{\mathbf{k}^2}{2m}-\mu$, $E_{\bk\uparrow,\downarrow}=E_\bk\mp h$ and $|u_{\bk}|^2,|v_{\bk}|^2=\frac{1}{2}(1\pm\frac{\xi_{\bk}}{E_{\bk}})$. Interestingly, the $q$BCS model has a built-in nature of population-imbalanced structure since the deformation parameters for each species must have the relation $q\leftrightarrow\frac{1}{q}$.
 When $q=1$ this model reduces to the ordinary BCS theory for population-imbalanced systems. Defining $x^+=(\tau^+,\bx)$ with $\tau^+=\tau+0^+$, the particle number densities for two species can be obtained from Green's functions by
\begin{align}\label{Number}
n_{\uparrow}=\frac{1}{qV}\int\dif^3\bx\; G_{\uparrow}(x,x^{+}),\quad
n_{\downarrow}=\frac{q}{V}\int\dif^3\bx\; G_{\downarrow}(x,x^{+}),
\end{align}
and the gap equation can be
deduced by either one of the anomalous Green's functions
\begin{align}
\Delta^{*}=-gF^{\dag}_{\uparrow\downarrow}(x^{+},x) =gqF^{\dag}_{\downarrow\uparrow}(x^{+},x).
\end{align}
Therefore, the equations of state are given by Eqs.(\ref{eud}) and Eqs.(\ref{edu})
\begin{align}\label{eofs}
n_{\uparrow}&= \frac{1}{q}\sum_{\bk}\Bigg[|u_{\bk}|^2f\left(E_{\bk\uparrow}-\frac{\ln q}{\beta}\right)+|v_{\bk}|^2f\left(-E_{\bk\downarrow}-\frac{\ln q}{\beta}\right)\Bigg], \nonumber \\
n_{\downarrow}&=q\sum_{\bk}\Bigg[|u_{\bk}|^2f\left(E_{\bk\downarrow}+\frac{\ln q}{\beta}\right)+|v_{\bk}|^2f\left(-E_{\bk\uparrow}+\frac{\ln q}{\beta}\right)\Bigg],\nonumber \\
-\frac{1}{g}&=\sum_{\bk}\frac{1-f\left(E_{\bk\uparrow}-\frac{\ln q}{\beta}\right)-f\left(E_{\bk\downarrow}+\frac{\ln q}{\beta}\right)}{2E_\bk}.
\end{align}
The quasiparticle energy dispersions are shifted by the deformation parameter as $E_{\bk\uparrow}-\frac{\ln q}{\beta}$, and $E_{\bk\downarrow}+\frac{\ln q}{\beta}$. Thus, the extra term $\frac{\ln q}{\beta}=T\ln q$ behaves like a shift of the chemical potentials.
Moreover, it can be found that the field algebra is invariant under the simultaneous interchanges $\uparrow\leftrightarrow\downarrow$ and $q\leftrightarrow\frac{1}{q}$. This symmetry is also reflected in the equations of motion and equations of state. Note under the interchange $\uparrow\leftrightarrow\downarrow$, $q\leftrightarrow\frac{1}{q}$, together with $h\leftrightarrow-h$ and $\Delta\leftrightarrow \frac{1}{\sqrt{q}}\Delta$, the equations (\ref{dEGAG}) and (\ref{mEGAG}) change to each other, so do $n_\uparrow$ and $n_\downarrow$ in Eqs.(\ref{eofs}). In numerics, the ground-state phase diagram also exhibits this symmetry\cite{OurJStat18}.

 The coupling constant $g$ in fact denotes a contact potential. This is not true in realistic systems since there must be a lower bound for the inter-particle distance, which further leads to an upper bound for quasiparticle momentum. Instead of introducing a momentum cutoff, the coupling constant can be renormalized by the $s$-wave scattering length $a$ of two-body scattering via
\begin{equation}\label{ren}
\frac{1}{g}=\frac{m}{4\pi a}-\int \frac{d^3\bk}{(2\pi)^3}\frac{1}{2\epsilon_\bk},
\end{equation}
where $\epsilon_\bk=\frac{\bk^2}{2m}$.
This regularization procedure of coupling constant must not be affected by the deformation of the quantization algebra for field operators since it is only a few-body effect and is obtained in quantum mechanics where $\psi_{\uparrow,\downarrow}$ is simply treated as a wave-function instead of an operator. The scattering length can be tuned by external magnetic field, and the associated two-body scattering problem has been well studied. Instead of $g$, in the following discussions, the dimension-less quantity $1/(k_Fa)$ with $k_F$ being the Fermi momentum to be defined later will be used to indicate the interacting strength.

\section{Numerical Analysis}               \label{sc2}
Now we try to visualize some interesting properties of the ultracold $q$-Fermi gas by numerical analysis. Here we need to point out that the $q$-Fermi gas theory is a mean-field approach without including other fluctuations since the present model is already complicated enough. We introduce the number polarization $p\equiv \delta n/n$ where $\delta n =n_\uparrow-n_\downarrow$ and $n=n_\uparrow+n_\downarrow$ are number difference density and total number density respectively. To normalize the physical quantities in which we are interested, $k_F$ satisfying $n=k_F^3/(3\pi^2)$, which is the Fermi momentum
of the ordinary unpolarized noninteracting Fermi gas with the same total number density as the
$q$-Fermi gas under studying,
is chosen as the unit for momentum. Accordingly, the unit of temperature is chosen as $k_BT_F=E_F=\hbar^2k_F^2/(2m)$, which is the noninteracting Fermi energy for $p=0$.

\begin{figure}[ht]
\centering
\includegraphics[width=2.2in, clip]{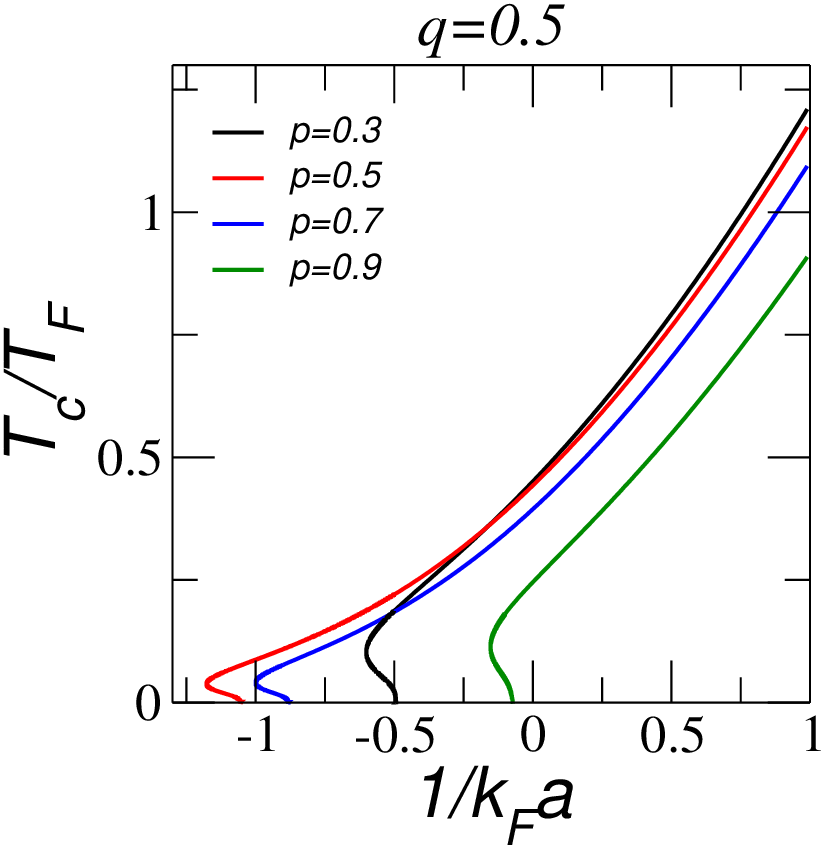}
\includegraphics[width=2.2in, clip]{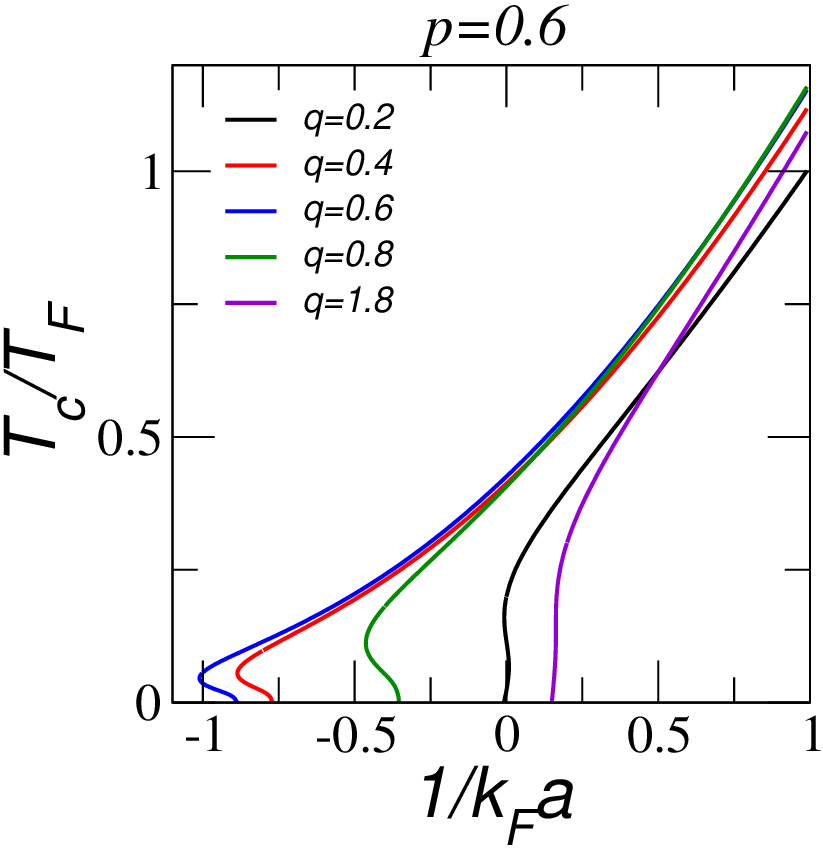}
 \caption{(Color online). $T_c$ as a function of $1/(k_Fa)$ of polarized $q$-Fermi gases with either fixed $q$ or $p$. In the left panel, $q$ is fixed to be $0.5$, while in the right panel $p$ is fixed to be $0.6$.}
 \label{fig.1}
\end{figure}

In the regime $1/(k_Fa)<0$ where the interaction is weak, the $q$-fermions are loosely paired at low temperatures. The condensed pairs are Cooper pairs in BCS theory, hence this regime is called BCS regime. In the regime $1/(k_Fa)>0$ where the interaction is strong, the pairs are tightly bound and behave like composite bosons since the system may be in a two-body bound states. These composite bosons are expected to form a condensate similar to BEC at low temperatures, thus the regime is called BEC regime. The point at which $1/(k_Fa)=0$ indicates the onset of two-body bound state and is then called unitary limit.
At zero temperature, the behaviors in the BCS regimes (or a superconducting model) has been studied in ref.\cite{OurJStat18}. It has been found that the breached pair structure\cite{WilczekPRL03} or Sarma phase\cite{Sarma} always exists even if $p=0$, which is qualitatively different from the ordinary Fermi gas. In other words, paring in the momentum space is present only when $k$ is below
\begin{equation}
k_1=\textrm{max}(0, \sqrt{2m(\mu-\sqrt{(h+T\ln q)^2-\frac{\Delta^2}{q}})})
\end{equation}
and above
\begin{equation}
k_2=\sqrt{2m(\mu+\sqrt{(h+T\ln q)^2-\frac{\Delta^2}{q}})}.
\end{equation}
This is a generalization of the Sarma phase.
We are interested in the effect of $q$ on ultracold Fermi gases in different regimes at finite temperatures.

Firstly, we need to find where the ordered phase with $\Delta>0$ exists at certain conditions. In Figure.\ref{fig.1}, we present the phase diagram by showing a plot of $T_c$ as a function of $1/(k_Fa)$ for various $p$ or $q$.
In the left panel, $q$ is fixed while $p$ varies, and $p$ is fixed while $q$ varies in the right panel.
The region where the pairing condensation exists is either below or on the right-hand-side of the $T_c$ lines. The other region is occupied by the phase of noninteracting $q$-Fermi gas. According to the figure, it can be found that in or close to the BCS regime, there may exist two $T_c$'s when $1/(k_Fa)$ is below a critical value, which is called intermediate-temperature superfluidity introduced in Ref.\cite{LombardoPRL00,Chien06,HeLYPRB06}. The reason can be briefly outlined as follows. Note pairings may exist even with relatively high energy $\xi_\bk>\xi_2\equiv k^2_2/(2m)-\mu$, which is due to the fact that some majority particles are ``excluded out'' from the ``normal'' majority species by Pauli principle to fill in states in the region $[k_1,k_2]$. Then the energy of the system is lowered by the condensation of these breached pairs. At low temperatures, the states within $[k_1,k_2]$ is almost completely filled and there is no room for pumped-out particles. Therefore, the system has to stay in the normal regime. Moreover, it can be found that in either panel, the superfluid regimes first expand and then contract as either $p$ or $q$ increases. This is due to the symmetry that the system possesses under the interchange $q\leftrightarrow\frac{1}{q}$ and $p\leftrightarrow-p$. The order parameter accordingly scales as $\Delta\leftrightarrow\frac{\Delta}{\sqrt{q}}$, which may bring changes to the ordered regimes.

\begin{figure}[ht]
\centering
\includegraphics[width=4.3in, clip]{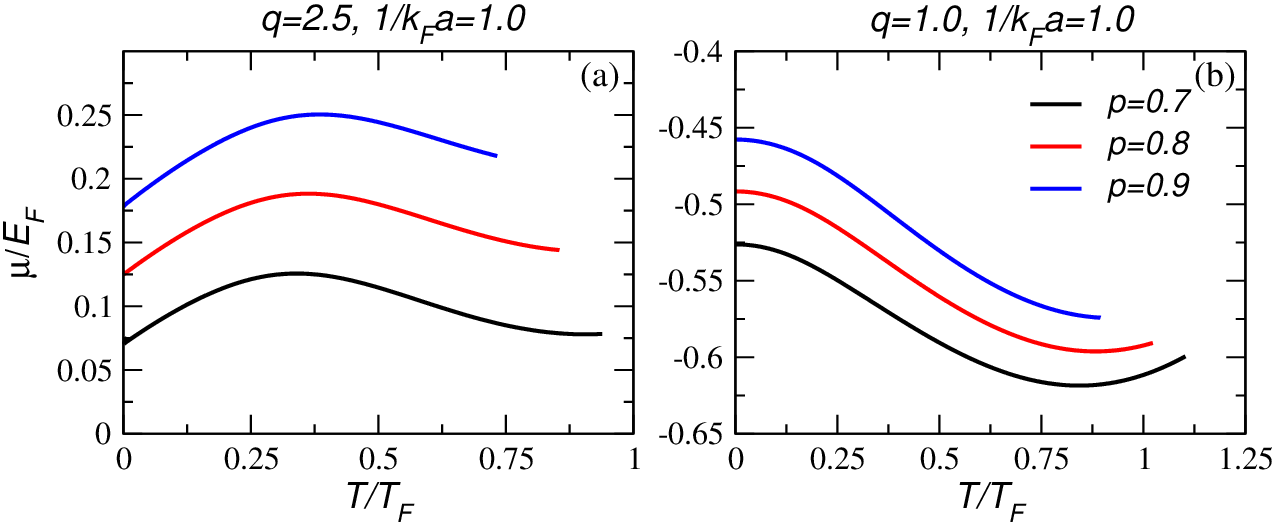}
 \caption{(Color online). Behaviors of chemical potentials of ultracold $q$-Fermi gases (left panel) and ordinary Fermi gases (right panel) in the BEC regime. The black, red, and blue solid lines correspond to polarized $q$-gases with $p=0.7,0.8$ and $0.9$ respectively. }
 \label{fig.2}
\end{figure}

Now we focused on the ordered regimes. Previously, we have pointed out that $\frac{\ln q}{\beta}=T\ln q$ behaves like a shift of chemical potentials.
At high temperatures where $\Delta\rightarrow 0$, $E_{\bk\uparrow,\downarrow}\rightarrow \xi_{\bk\uparrow,\downarrow}$, the quasiparticle energy dispersions approach $\xi_{\bk\uparrow,\downarrow}\mp \frac{\ln q}{\beta}$. By choosing suitable deformation parameters at certain temperatures, one may expect that the signs of the original chemical potentials are flipped, which may change the essential properties of the associated quasiparticles. If $q>1$ and $p>0$, it can be seen that the quasiparticle energy is shifted by $\pm T\ln q$ for minority and majority respectively. The left panel of Fig.\ref{fig.2}
 shows how the chemical potential of the $q$-gas with $q=2.5$ changes as the temperature increases (all curves stop at $T_c$) when $1/(k_Fa)=1.0$ which is in the BEC regime
 for ordinary Fermi gases with $q=1.0$. Here the black, red, and blue solid lines correspond to polarized $q$-gases with $p=0.7$, $0.8$ and $0.9$. It can be found that all chemical potentials become positive, while ordinary bosons with $q=1.0$ always have negative chemical potentials. As a comparison, in the right panel we show the corresponding behaviors of chemical potentials of ordinary Fermi gases, i.e. $q=1$ with the same population imbalances. This is a quite interesting phenomenon that the composite bosons behave like fermions under the tuning of the deformation parameter, which is never found before.

\begin{figure}[ht]
\centering
\includegraphics[width=4.3in, clip]{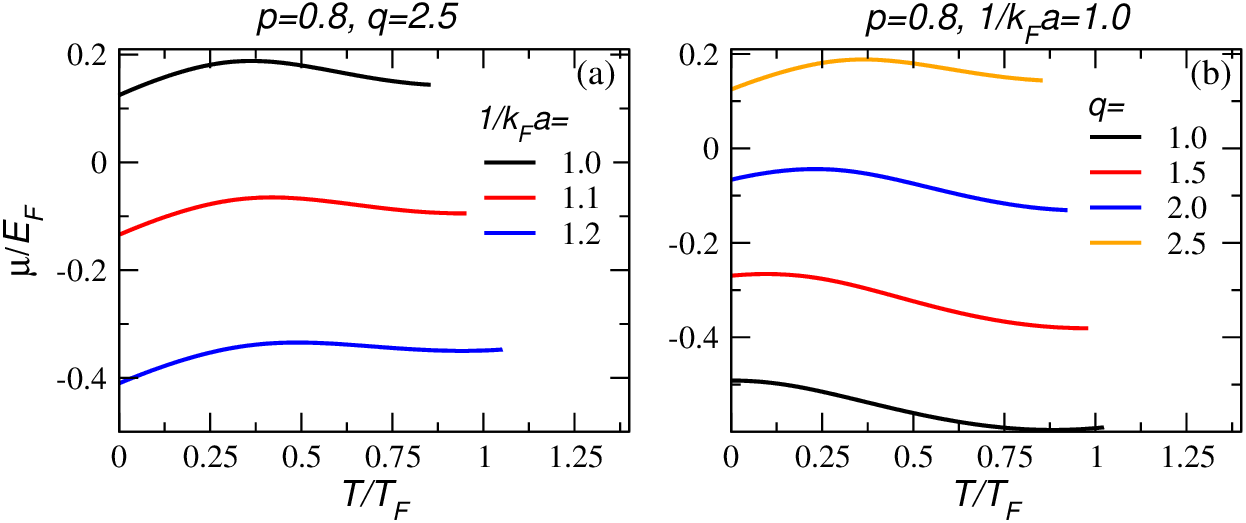}
 \caption{(Color online). The ``Fermi-Bose'' crossover of the $q$-gas. In the left panel, the chemical potential becomes negative again in the deeper BEC regime, i.e. with larger $1/k_Fa$. Here the black, red and blue solid lines correspond to the situations with $1/k_Fa=1.0,1.1$ and $1.2$ respectively. In the right panel, we show how the chemical potential gradually becomes negative by the tuning of $q$. Here the black, red, blue and orange solid lines correspond to the situations with $q=1.0,1.5,2.0$ and $2.5$ respectively.  }
 \label{fig.3}
\end{figure}

In Figure.\ref{fig.3}, we show the ``Bose-Fermi'' crossover of the $q$-gas by tuning various parameters. In panel (a) (left panel), $q$ is fixed as 2.5. We have shown in Figure.\ref{fig.2} that the chemical potential becomes positive when $1/k_Fa=1.0$, i.e. the $q$-gas behaves like a fermionic condensate. But in slightly deeper BEC regime, for example, $1/k_Fa=1.1,1.2$, the chemical potential quickly becomes negative again.
Hence the original BCS-BEC crossover is strongly effected by the tuning of the deformation parameter.
If we want the system to become ``fermionic" again, a larger deformation parameter $q$ is expected. In panel (b) (right panel), we show how the chemical potential of a $q$-gas gradually becomes positive as $q$ increases, which might be referred to as the $q$-induced ``Bose-Fermi'' crossover of the $q$-gas.

\begin{figure}[ht]
\centering
\includegraphics[width=2.3in, clip]{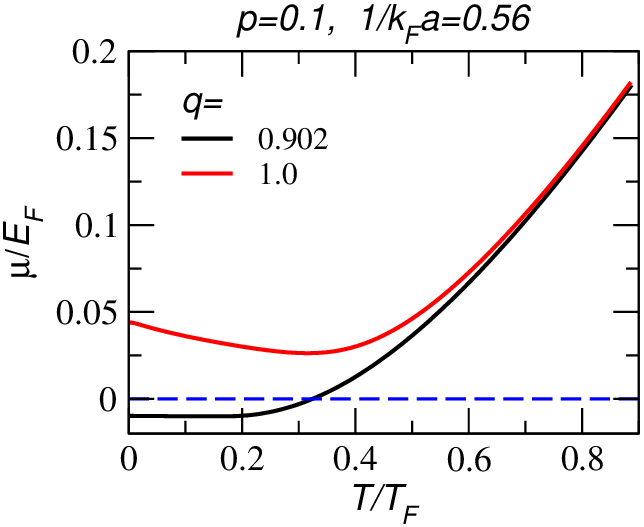}
 \caption{(Color online). Behaviors of chemical potentials of ultracold $q$-Fermi gases in the fermionic regimes. The black and red solid lines correspond to the situations with $q=0.902,1.0$ respectively.  }
 \label{fig.4}
\end{figure}

One might also be interested in the converse situation whether the chemical potential of a $q$-gas can change from positive (fermionic) to negative (bosonic) by tuning the deformation parameter $q$? Naively, it seems this can be achieved by choosing $0<q<1$ since $T\ln q$ now gives a negative contribution. However, as we pointed out previously, the system possesses a symmetry under the interchange $q\leftrightarrow\frac{1}{q}$ and $p\leftrightarrow-p$. Thus, the possible choices of parameters are highly constrained. Accordingly, what we find is a very ``weak'' signature as shown in Figure.\ref{fig.4}. When $p=0.1$ and $1/k_Fa=0.56$, the chemical potential at low temperatures indeed becomes negative (bosonic) when $q=0.902$. For the ordinary Fermi gas with $q=1.0$ at low temperatures, the chemical potential is positive but close to zero. It is known that the chemical potential of an ordinary Fermi gas changes sign at around $1/k_Fa\simeq 0.57$. Thus, ordinary Fermi-gases with $1/k_Fa\lesssim0.56$ are referred to as being already in the very weak fermionic side, and it is easy to change the sign of their chemical potential by tuning $q$ there. Nevertheless, the sign-changing signature of the chemical potential denotes an essential change of the physical properties.
 Hence, this signature of the $q$-gas can be safely recognized as a weak signal for the ``Fermi-Bose'' crossover from the ordinary Fermi gas to the $q$-Bose gas.

\section{Conclusion}
To summarize, we construct a microscopic theory for an interacting gas which obeys $q$-deformed Fermi statistics by incooperating with a $q$-deformed algebra. This ultracold quantum gas, which may be referred to as a $q$-gas, presents some interesting features. In the ordinary BEC regime ($1/k_Fa=1.0$), the system exhibits fermionic features, i.e. its chemical potential becomes positive, as a certain deformation parameter is chosen. In contrast, we also find a converse situation in which the originally fermionic system exhibits bosonic features, though it might be a weak signal.
\begin{acknowledgments}
This work was supported by the National Natural Science Foundation of China (Grant No. 11674051)
\end{acknowledgments}

\end{document}